\documentstyle[aps,epsf]{revtex}
\tightenlines
\newtheorem{theorem}{Theorem}

\newtheorem{criterion}[theorem]{Criterion}

\begin{document}
\title{Peculiarities of the Weyl - Wigner - Moyal formalism for scalar charged
particles\footnote{J. Phys. A: Math. Gen. 34 (2001) 4323 – 4339,
www.iop.org/Journals/ja\\ (c) 2001 IOP Publishing Ltd}}
\author{B.I. Lev$^{1,2}$\footnote{%
Electronic addresses: lev@iop.kiev.ua, blev@i.kiev.ua}}
\author{A.A. Semenov$^1$\footnote{%
Electronic address: sem@iop.kiev.ua}}
\author{C.V. Usenko$^2$\footnote{%
Electronic addresses: usenko@phys.univ.kiev.ua,
usenko@ups.kiev.ua}}
\address{$^1$Institute of Physics, National Academy of Science of Ukraine.\\46
Nauky pr., Kiev 03028, Ukraine \\
$^2$Physics Department, Taras Shevchenko Kiev University.\\6
Academician Glushkov pr., Kiev 03127, Ukraine}
\date{21 February 2001 }
\maketitle

\begin{abstract}
A description of scalar charged particles, based on the Feshbach -
Villars formalism, is proposed. Particles are described by an
object that is a Wigner function in usual coordinates and momenta
and a density matrix in the charge variable. It is possible to
introduce the usual Wigner function for a large class of
dynamical variables. Such an approach explicitly contains  a
measuring device frame. From our point of view it corresponds to
the Copenhagen interpretation of quantum mechanics. It is shown
how physical properties of such particles depend on the
definition of the coordinate operator. The evolution equation for
the Wigner function of a single-charge state in the classical
limit coincides with the Liouville equation. Localization
peculiarities manifest themselves in specific constraints on
possible initial conditions.
\end{abstract}

\pacs{0365C}

\section{Introduction}

\label{Introduction}

The question regarding the nature of the wavefunction has its
origin in the early years of quantum mechanics. For a long time it
has been considered as a philosophical rather then a physical
question \cite{b1,b2,b3}. However, it is now very real because of
the recent theoretical and experimental progress in quantum
information \cite{b2,b4,b5}.

One of the significant points in understanding the nature of the
wave function is the Einstein - Podolsky - Rosen paradox and the
existence of quantum correlations related to it. Since 1980, such
specific behavior of quantum systems has been confirmed many
times in experiments \cite{b6}. It is very important that such
correlations ''spread'' in space instantly. Nevertheless, if one
adheres to the Copenhagen interpretation, there is no violation of
causality principle.

However, due to the fact that collapse of the wavefunction takes
place at a given instant (or maybe in a small time interval) in
the whole space, one can speak of violation of causality principle
and a conflict between quantum mechanics and special relativity
\cite{b7,b8} in some interpretations of quantum mechanics.
Following the idea of Bell and Eberhard  that a consistent
description should contain a preferred frame, relativistic
classical and quantum mechanics is built in \cite{b9,b10} in such
a way that the relativity principle is generalized and the
causality is not violated. A basic assumption of this theory is
that transformation from one frame to another is realized by
operators which are isomorphic to the Lorentz group and depend on
a certain four-vector (as parameter) in such a way that the
instant time hyper-plain is invariant. This vector is interpreted
as a relative to an observer four-velocity of the preferred frame.
Under this assumption the postulate about the light speed
constancy is changed to the postulate about the constancy of the
average speed of light on closed path.

One of the remarkable peculiarities of this theory is in the
existence of a well-defined position operator \cite{b10} which
coincides with the Newton - Wigner position operator
\cite{b11,b12} in the preferred frame. It means that in this
approach measurement of position does not create a particle -
antiparticle couple since the odd part responsible for appearance
of a different charge state superposition is absent. Hence, there
exists a supposition that localization of relativistic particles
can bear information about presence of the preferred frame in the
Universe.

The above arguments show the fundamental role that determination
of the position operator structure can play. Moreover, the
question of whether it is a one-particle, or, in principle, a
many-particle operator, is of specific interest as well.
Therefore, it is very important to find situations where the odd
part of the position operator could manifest itself. It would be
possible to realize such observation on strongly localized (near
the Compton wavelength) states of a separate particle. However,
because such states are a problem for laboratory experiments, it
is of great interest how this peculiarity manifests itself in a
many-particle system.

The Weyl - Wigner - Moyal (WWM) formalism is a convenient method
to describe both one-particle and many-particle quantum systems
in non-relativistic theory \cite{b13,b14,b15,b16}. Nevertheless,
attempts to generalize it for the relativistic case lead to a
number of problems. The first problem is that the Weyl rule
\cite{b17} does not include time as a dynamical variable, and the
scalar product in the Hilbert space of states is formulated for
functions square integrable not over the whole space-time but in
the three-dimensional space or in a space-like hyper-surface
only. In [18] this problem was resolved by generalization of the
spatial integration over the whole space-time without Weyl rule
application. The WWM formalism in the framework of the stochastic
formulation of quantum mechanics, where the scalar product is
formulated for square integrable functions in the whole
space-time, also leads to Lorenz-invariant expressions \cite{b19}.

The matrix-valued Wigner function formalism has been developed
for the general case of many component equations and, in
particular, for $1/2$ spin particles on the basis of the usual
Weyl rule \cite{b20,b21}. Certainly, such equations are not Lorenz
invariant, though average values coincide with their analogues in
the usual approach.

The next essential problem in relativistic WWM formalism is in
absence of a well-defined position operator. Unlike the works
mentioned above, where the Wigner function is determined by means
of the usual position operator, in \cite {b22,b23} a formalism
with the Newton - Wigner position operator is developed. This
approach is not Lorenz invariant either, and respective results
differ from the standard ones. However, they can be related to
\cite{b10}, where the consistent definition of the position
operator is possible.

The goals of this paper are in formulating the WWM relativistic
formalism for scalar charged particles under the approach of
\cite{b20} and finding a set of specific peculiarities in
relativistic quantum system behavior, that are related to the
non-trivial structure of the position operator. It turns out that
values of some observables depend directly on the position
operator definition, that, as mentioned above, can be a
consequence of existence of the preferred frame in the Universe.

In Sec.\ref{II} we introduce the Weyl rule for matrix-valued
observables in the case of scalar charged particles and discuss
peculiarities of correspondence between classical and quantum
dynamical algebras. Sec.\ref {III} is devoted to the
matrix-valued Wigner function and quantum Liouville equation.
Here we also discuss the absence of Lorenz invariance of this
approach, and how it can be related to the Copenhagen
interpretation of quantum mechanics. In Sec.\ref{IV} we consider
how the usual Weyl rule turns into the Feshbach - Villars
representation. Among the whole set of dynamical variables, we
separate a special class of observables with the Weyl symbols
independent of charge variable (charge-invariant variables). They
have several remarkable properties. In particular, we find the
relationship between their even and odd parts. The usual (not
matrix-valued) Wigner function can be introduced for such
observables. It is considered in Sec.\ref {V} (for a brief
abstract of this approach see \cite{b24}). This object includes
four components: one corresponds to a particle, second one
corresponds to an antiparticle (even part of Wigner function), and
two more are interference terms (odd part of Wigner function).
The evolution equation for the odd part becomes zero in the
classical limit, and the equation for the even part coincides with
its analogue in the Newton - Wigner position operator approach
\cite{b22,b23}. The difference reveals itself in peculiarities of
constraints on initial conditions that are considered in
Sec.\ref{VI}.

\section{Weyl rule and specific properties of dynamical algebra for scalar
charged particles}

\label{II}

In this and the following Sections we apply the methods developed
in \cite {b20,b21} to the Klein - Gordon equation that is written
using the Feshbach - Villars formalism \cite{b12}. This makes it
possible to take the charge variable into account explicitly.
With account of the fact that the Hilbert space of states for
scalar charged particles has an indefinite metric, we have to
distinguish between covariant and contravariant basis vectors and
coordinates for subspace corresponding to the charge variable.
Hence, arbitrary state is expanded on basis vectors in the
following way:
\begin{equation}
\left| \Psi \right\rangle =\sum\limits_{\alpha =\pm 1}{\left| {\Psi ^{\alpha
}}\right\rangle \otimes \left| {e_{\alpha }}\right\rangle },  \label{f1}
\end{equation}
where
\begin{equation}
\left| {\Psi ^{\alpha }}\right\rangle =\sum\limits_{s}{\Psi ^{\alpha
}(s)\left| s\right\rangle }.  \label{f2}
\end{equation}
Here $s=q,p,...$ is an arbitrary dynamical variable and the
summarizing symbol can be interpreted as an integration with
respect to $s$. Greek indices take values $\pm 1$. $\tau
_{3}{}_{\alpha \beta }$ plays the role of metric tensor (see
\cite{b12}). In notations like $\delta _\alpha {}^{\beta } $,
$\tau _{1_{\alpha }\beta }$, $\tau _{2\alpha \beta }$, $\tau
_{3\alpha \beta }$, $\tau _{3}{}_\alpha {}^\beta $,..., symbols
$\delta $, $\tau _{i}$ do not correspond to specific operators
and mean only values of the corresponding matrix elements. There
are some peculiarities in notations of operators as well.
$\hat{A}$ is a full operator that acts on all dynamical
variables. $\check{A}$ is an operator that acts on the charge
variables only (or, in other words, it is a $c$-numeric matrix).

For a consistent development of the WWM formalism we shall
formulate the Weyl rule bringing into correspondence
matrix-valued classical variables (Weyl symbols) to
quantum-mechanical operators. Here we should take into account
that in our case classical variables are operators acting on the
charge variable. However, because the operator of
quasi-probability density in the coordinate representation is
proportional to the identity matrix in charge space, the
succession of this operator and a matrix-valued Weyl symbol does
not matter. Choosing it in an arbitrary way, one can write the
Weyl rule as follows:
\begin{equation}
\hat A_\alpha {}^\beta =\sum\limits_{\gamma =\pm 1}{\int\limits_{-\infty
}^{+\infty }{A_\gamma {}^\beta (p,q)\hat W_\alpha {}^\gamma (p,q)dpdq}},
\label{f3}
\end{equation}
where $A_\alpha {}^\beta (p,q)$ is the matrix-valued Weyl symbol, $%
\hat A_\alpha {}^\beta $ stands for corresponding quantum-mechanical
operator, $\hat W_\alpha {}^\beta (p,q)$ is the operator of
quasi-probability density that is the Fourier transform of the displacement
operator. Following \cite{b16} one can obtain the expansion of $\hat W%
_\alpha {}^\beta (p,q)$ on eigenvectors of the position operator:
\begin{equation}
\hat W_\alpha {}^\beta (p,q)=\frac 1{{(2\pi \hbar )^d}}\int\limits_{-\infty
}^{+\infty }{\left| {q+\frac Q2}\right\rangle \delta _\alpha {}^\beta \exp\left[%
\frac i\hbar Qp\right]dQ\left\langle {q-\frac Q2}\right| },
\label{f4}
\end{equation}
where $d=3$ is dimensionality of the physical space.

Substituting (\ref{f4}) to (\ref{f3}) one can find the expression
for an arbitrary operator expansion:
\begin{equation}
\hat{A}_{\alpha }{}^{\beta }=\frac{1}{{(2\pi \hbar )^{d}}}%
\int\limits_{-\infty }^{+\infty }{\left| {q+\frac{Q}{2}}\right\rangle
A_{\alpha }{}^{\beta }(p,q)}\exp\left[\frac{i}{\hbar }Qp\right]dQdpdq\left\langle {q-%
\frac{Q}{2}}\right| .  \label{f5}
\end{equation}
Similarly, such expression can be written by expending the
operator of quasi-probability density on the eigenvectors of the
momentum operator:
\begin{equation}
\hat{A}_{\alpha }{}^{\beta }=\frac{1}{{(2\pi \hbar )^{d}}}%
\int\limits_{-\infty }^{+\infty }{\left| {p+\frac{P}{2}}\right\rangle
A_{\alpha }{}^{\beta }(p,q)\exp\left[-\frac{i}{\hbar }Pq\right]dPdpdq\left\langle {p-%
\frac{P}{2}}\right| }.  \label{f6}
\end{equation}
Matrix elements of this operator can be found in the form:
\begin{equation}
\left\langle {q_{1}}\right| \hat{A}_{\alpha }{}^{\beta }\left| {q_{2}}%
\right\rangle =\frac{1}{{(2\pi \hbar )^{d}}}\int\limits_{-\infty }^{+\infty }%
{A_{\alpha }{}^{\beta }(p,\frac{1}{2}(q_{1}+q_{2}))\exp\left[\frac{i}{\hbar }%
(q_{1}-q_{2})p\right]dp}.  \label{f7}
\end{equation}
By changing variables in this expression in a standard way and
performing Fourier transformation one can obtain the expression
that reconstructs matrix-valued Weyl symbol through operator:
\begin{equation}
A_{\alpha }{}^{\beta }(p,q)=\int\limits_{-\infty }^{+\infty }{\left\langle {%
q+\frac{Q}{2}}\right| \hat{A}_{\alpha }{}^{\beta }\left| {q-\frac{Q}{2}}%
\right\rangle \exp\left[-\frac{i}{\hbar }Qp\right]dQ}.  \label{f8}
\end{equation}

In \cite{b22} the integral form of the Klein - Gordon equation
with pseudo-differential symbols in the Feshbach - Villars
representation (where Hamiltonian matrix has a diagonal form) is
obtained. In fact, it means that the Newton - Wigner coordinate
is used there instead of the usual coordinate. Here we obtain
another integral form of the Klein - Gordon equation using the
usual position operator:
\begin{equation}
i\hbar \partial _{t}\Psi _{\alpha }(x)=\frac{1}{{(2\pi \hbar )^{d}}}%
\sum\limits_{\beta =\pm 1}{\int\limits_{-\infty }^{+\infty
}{H^{\beta} {}_{\alpha
}(p,\frac{1}{2}(x+y))\exp\left[\frac{i}{\hbar }(x-y)p\right]\Psi
_{\beta }(y)dpdy}},  \label{f9}
\end{equation}
where
\begin{equation}
H^{\beta}{}_{\alpha }(p,x)=(\tau _{3}{}^{\beta }{}_{\alpha }+i\tau
_{2}{}^{\beta} {}_{\alpha })\frac{{(\vec{p}-e\vec{A}(x))^{2}}}{{2m}}+\tau
_{3}{}^{\beta} {}_{\alpha }mc^{2}+e\varphi (x)\delta ^{\beta }{} _{\alpha}
\label{f10}
\end{equation}
is the matrix-valued Weyl symbol of the Hamiltonian in the general
case (for a particle in electromagnetic field), which depends on
both coordinate and momentum. Unlike \cite{b22} this equation
takes into account that the position operator mixes states with
different charge signs.

Until now,consideration of the matrix-valued WWM formalism has
been very similar to the usual one. Considerable differences
appear in determining the matrix-valued Moyal bracket and in the
classical limit. To consider this question one should find the
matrix-valued Weyl symbol from the product of two operators in a
standard way.

Let $\check{A}(p,q)$ and $\check{B}(p,q)$ be matrix-valued Weyl symbols of
two operators $\hat{A}$ and $\hat{B}$ respectively. Let us introduce
operators $\hat{C}_{1},\hat{C}_{2},\hat{C}$ as follows:
\begin{equation}
\begin{array}{l}
\hat{C}_{1}=\hat{A}\hat{B} \\
\hat{C}_{2}=\hat{B}\hat{A} \\
\hat{C}=[\hat{A},\hat{B}]=\hat{C}_{1}-\hat{C}_{2}
\end{array}
.  \label{f11}
\end{equation}
Then, using (\ref{f8}) and consideration like that used in
\cite{b15}, one can obtain the matrix-valued Weyl symbol of the
operator $\hat{C}_{1}$:
\begin{equation}
\check{C}_{1}(p,q)=\check{A}(p,q)\exp\left[\frac{{i\hbar }}{2}(\overleftarrow{%
\partial }_{q}\overrightarrow{\partial }_{p}-\overleftarrow{\partial }_{p}%
\overrightarrow{\partial }_{q})\right]\check{B}(p,q).  \label{f12}
\end{equation}
For the matrix-valued Weyl symbol of the operator $\hat{C}_{2}$ one can find
a similar expression. Therefore, the matrix-valued Moyal bracket can be
determined in the following form:
\begin{equation}
\begin{array}{l}
\{\check{A}(p,q),\check{B}(p,q)\}_{M}=\frac{1}{{i\hbar }}\check{C}(p,q)= \\
=\frac{1}{{i\hbar }}\left( {\check{A}(p,q)\exp\left[\frac{{i\hbar }}{2}(\overleftarrow{%
\partial }_{q}\overrightarrow{\partial }_{p}-\overleftarrow{\partial }_{p}%
\overrightarrow{\partial }_{q})\right]\check{B}(p,q)-\check{B}%
(p,q)\exp\left[\frac{{i\hbar }}{2}(\overleftarrow{%
\partial }_{q}\overrightarrow{\partial }_{p}-\overleftarrow{\partial }_{p}%
\overrightarrow{\partial }_{q})\right]%
\check{A}(p,q)}\right)
\end{array}
.  \label{f13}
\end{equation}
Unlike the usual case, matrix-valued Weyl symbols do not
commutate, so it is not possible to represent such a Moyal
bracket as a sinus from the operator of the Poisson bracket. This
fact results in the classical limit in some peculiarities. In the
general case, one can express the classical limit of the
matrix-valued Moyal bracket via the matrix-valued Poisson bracket
\cite{b21} and the commutator of the matrix-valued Weyl symbols:

\begin{equation}
\mathrel{\mathop{\lim}\limits_{\hbar \rightarrow 0}}%
{{\{\check{A},\check{B}\}}_{M}}=\frac{1}{{i\hbar }}[\check{A},\check{B}]+%
\frac{1}{2}\left( {\{\check{A},\check{B} \}_{P}-\{\check{B},\check{A}\}_{P}}%
\right) .  \label{f14}
\end{equation}

If the matrix-valued Weyl symbols (for example position and
Hamiltonian) commutate, the matrix-valued Moyal bracket is in fact
the usual one and coincides with the Poisson bracket in the
classical limit. If they do not commutate with each other, three
cases are possible. Let $[\check{A} ,\check{B} ] = O\left( {\hbar
^\nu } \right)$ when $\hbar \to 0 $. If $\nu < 1$, the classical
limit does not exist for such a couple of the matrix-valued Weyl
symbols, since the value of the matrix-valued Moyal bracket increases infinitely. If $%
\nu > 1$, it becomes zero in the classical limit. If $\nu = 1$,
the classical limit exists. Here we can take as an example the
Newton -Wigner position operator \cite{b11,b12}. Part of its
matrix-valued Weyl symbol,not commutating with the Hamiltonian, is
proportional to Planck's constant, thus in this case there exists
a well-defined classical limit.

\section{Matrix-valued Wigner function and quantum Liouville equation}

\label{III}

Using (\ref{f3}), one can find the average value of an arbitrary operator $%
\hat{A}$ expressed via its matrix-valued Weyl symbol:
\begin{equation}
\bar{A}=\sum\limits_{\alpha ,\beta ,\gamma =\pm 1}{\int\limits_{-\infty
}^{+\infty }{A_{\gamma }{}^{\beta }(p,q)\left\langle {\Psi _{\beta }}\right|
\hat{W}_{\alpha }{}^{\gamma }(p,q)\left| {\Psi ^{\alpha }}\right\rangle dpdq}%
}.  \label{f15}
\end{equation}
Then, introducing the matrix-valued Wigner function
\begin{equation}
W_{\beta }{}^{\gamma }(p,q)=\sum\limits_{\alpha =\pm 1}{\left\langle {\Psi
_{\beta }}\right| \hat{W}_{\alpha }{}^{\gamma }(p,q)\left| {\Psi ^{\alpha }}%
\right\rangle },  \label{f16}
\end{equation}
this expression can be written in a simpler form:
\begin{equation}
\bar{A}=tr\int\limits_{-\infty }^{+\infty }{\check{A}(p,q)\check{W}(p,q)dpdq}%
.  \label{f17}
\end{equation}

Substitution of the operator of quasi-probability density
expansion (\ref{f4}) into (\ref{f16}) leads to the expression for
the matrix-valued Wigner function in the coordinate
representation. As a result we obtain a formula that coincides
with the definition given in \cite{b20}:
\begin{equation}
W_{\alpha }{}^{\beta }(p,q)=\frac{1}{{(2\pi \hbar )^{d}}}\int\limits_{-%
\infty }^{+\infty }{\Psi _{\alpha }^{\ast }(q+\frac{Q}{2})\Psi ^{\beta }(q-%
\frac{Q}{2})\exp\left[\frac{i}{\hbar }Qp\right]dQ}.  \label{f18}
\end{equation}

Next, let us describe a method to obtain evolution equation for
the matrix-valued Wigner function of a scalar charged particle.
To do this one shall differentiate expression (\ref{f18}) with
respect to time. The value of the wavefunction derivative is
taken from the integral form of the Klein - Gordon equation
(\ref{f9}). After standard transformations, similar to the usual
WWM formalism, we obtain the quantum Liouville equation
\begin{equation}
\partial _t{\check W}=\{\check H,\check W\}_M,  \label{f19}
\end{equation}
where the matrix-valued Moyal bracket is determined by expression
(\ref{f13}).

At first sight, this equation has a considerable defect in
comparison to similar expressions in \cite{b18,b19}; namely, it is
not Lorentz invariant. For an interpretation of this fact, we have
to take into account arguments from the measurement theory. The
fact is that average values calculated in this approach coincide
with ones in the usual (Schr\"{o}dinger) representation of quantum
mechanics that is Lorentz invariant. Nevertheless, unlike the
stochastic formulation of quantum mechanics \cite{b25}, the scalar
product is determined here with functions that are square
integrable in a certain space-like hyper-surface, rather than over
the whole space-time. Moreover, the value of scalar product does
not depend on its choice \cite{b26}. One can relate this
hyper-surface to the measurement device frame. In other words, the
wavefunction collapse occurs in a frame in which equation
(\ref{f19}) is written. The absence of Lorentz invariance is a
consequence of the fact that the Weyl rule does not include time
as an independent dynamical variable.

\begin{figure}\centerline{\epsfysize=10.0cm \epsfbox{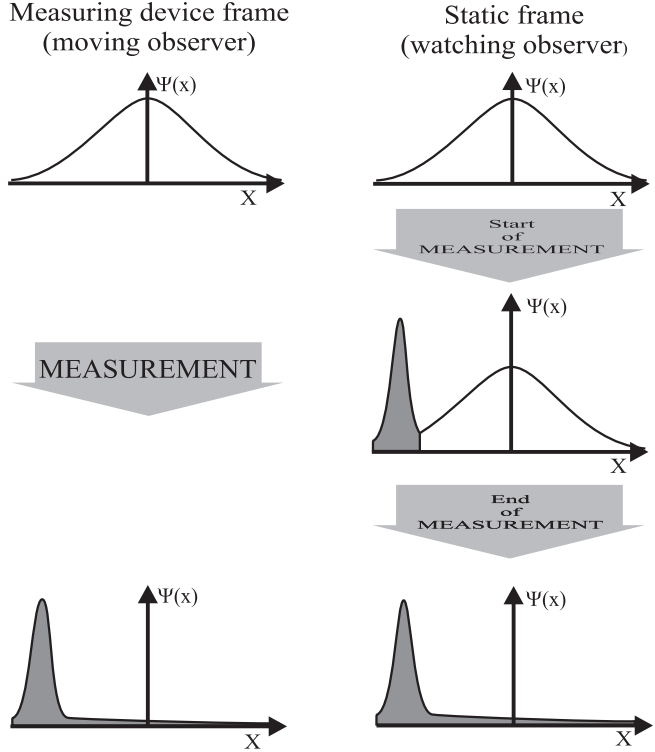}}
\caption{ The process of the wavefunction collapse in a moving
frame from points of view of an observer in this frame and a
static observer (watching observer). For the latter this process
is a certain moving front. At a certain instant a state exists
where at one side (right) there is the wavefunction before the
measurement and on the other side (left) there is the
wavefunction after the measurement.} \label{fig1}
\end{figure}

In \cite{b27} the process of relativistic measurement is
considered. In that work the point of view is expressed, that the
wavefunction (and as result the Wigner function, we notice) has
no objective value and does not covariantly transform  when there
are classical interventions. Here we note that in principle
equation (\ref{f19}) can be written with four-dimensional Lorentz
invariant symbols only, but to do this we have to incorporate in
the theory a certain time-like unit vector in a way similar to
Tomonaga-Schwinger approach to quantum field theory
\cite{b271,b272}. It is the four-velocity of the frame where the
wavefunction collapse occurs (the measuring device frame)
relative to the second static observer (watching observer). This
frame has principally another sense to that of the preferred
frame in \cite{b7,b8,b9,b10}. In that approach \cite{b9,b10}
collapse of the wavefunction has to take place in all frames in
the whole space (because of the instant time hyper-surface being
invariant). In our approach this process obeys the relativity of
simultaneity. If it were possible to observe the wave function
directly, such a hypothetical watching observer would see the
wavefunction collapse as a certain moving front. As a result, at
a certain instant, in a static frame (attached to the watching
observer), a state with a part of the wavefunction before
measurement on one hand and a part of the wavefunction after
measurement on the other hand, is realized (Fig.\ref{fig1}). But
perhaps it is impossible to propose even gedanken experiment
without use of a superluminar signal to interpret the results in
favor of one of the approaches.

Hence we find to be very important and of fundamental value the
fact that quantum mechanics in the Wigner representation
necessarily includes the four-velocity of the measuring device
frame in explicit form.

\section{Transformation to the Feshbach - Villars representation}

\label{IV}

In the Feshbach - Villars representation, Hamiltonian matrix has
a diagonal form and indices in the charge space correspond to the
particle and antiparticle \cite{b12}. Therefore, it is convenient
to distinguish between solutions with different charge signs and
to give them explicit physical sense. Furthermore, in this
approach an influence of the odd part of the position operator on
values of some physical variables is more evident.

Operator $\hat A_\alpha {}^\beta $transforms to the Feshbach -
Villars representation by the following formula:
\begin{equation}
\hat A^{FV}{}_\alpha {}^\beta =\sum\limits_{\gamma ,\delta =\pm 1}{\hat U%
_\gamma {}^\beta \hat A_\delta {}^\gamma \hat U^{-1}{}_\alpha {}^\delta },
\label{f20}
\end{equation}
where the transformation matrix has the form:
\begin{equation}
\hat U_\alpha {}^\beta (\hat p)=\frac 1{{2\sqrt{mc^2E(\hat p)}}}\left[ {(E(%
\hat p)+mc^2)\delta _\alpha {}^\beta +(E(\hat p)-mc^2)\tau _1{}_\alpha
{}^\beta }\right] .  \label{f21}
\end{equation}
Here and below
\begin{equation}
E(p)=\left(m^2c^4+c^2p^2\right)^\frac{1}{2}  \label{f22}
\end{equation}
is the energy of a relativistic free particle.

Next, we apply (\ref{f20}) to the Weyl rule in the form
(\ref{f6}). As result, we obtain the expression that brings into
correspondence the matrix-valued Weyl symbol to the operator in
the Feshbach-Villars representation:
\begin{equation}
\hat A^{FV}{}_\alpha {}^\beta =\frac 1{{(2\pi \hbar )^d}}\int\limits_{-%
\infty }^{+\infty }{\left| {p+\frac P2}\right\rangle
}\sum\limits_{\gamma ,\delta =\pm 1}{U_\gamma {}^\beta (p+\frac
P2)A_\delta {}^\gamma (p,q)U^{-1}{}_\alpha {}^\delta (p-\frac
P2)\exp\left[-\frac i\hbar Pq\right]dPdpdq\left\langle {p-\frac
P2}\right| }. \label{f23}
\end{equation}
In general, it is rather difficult to interpret this expression
since there is a complicated dependence on the integration
variable under the integral sign. Here we do not go beyond the
simplest case.

Let the matrix-valued Weyl symbol be proportional to the identity matrix:
\begin{equation}
A_\alpha {}^\beta (p,q)=A(p,q)\delta _\alpha {}^\beta .  \label{f24}
\end{equation}
In principle, one can say that such symbols do not depend on the
charge variable, so that the class of dynamical variables, which
corresponds to those, we denominate here, to be more brief, as a
class of charge-invariant variables. Most of the dynamical
variables that we consider in relativistic (non-quantum)
mechanics belong to this class. The reason for this is the absence
of a dependence on the charge variable in classical mechanics.
Hence, the question about the classical limit is for such
variables especially interesting.

The Weyl rule for charge-invariant variables in the Feshbach - Villars
representation has the form:
\begin{equation}
\hat A^{FV}{}_\alpha {}^\beta =\frac 1{{(2\pi \hbar )^d}}\int\limits_{-%
\infty }^{+\infty }{\left| {p+\frac P2}\right\rangle A(p,q)R_\alpha {}^\beta
(p+\frac P2,p-\frac P2)\exp\left[-\frac i\hbar Pq\right]dPdpdq\left\langle {p-\frac P2}%
\right| }.  \label{f25}
\end{equation}
Unlike \cite{b23} and non-relativistic case there is a matrix-valued
variable here:
\begin{equation}
R_\alpha {}^\beta (p_1,p_2)=\varepsilon (p_1,p_2)\delta _\alpha {}^\beta
+\chi (p_1,p_2)\tau _1{}_\alpha {}^\beta .  \label{f26}
\end{equation}
It contains even and odd parts and is expressed via the energy of
a free particle (\ref{f22}):
\begin{equation}
\begin{tabular}{l}
$\varepsilon (p_1,p_2)=\frac{{E(p_1)+E(p_2)}}{{2\sqrt{E(p_1)E(p_2)}}}$ \\
$\chi (p_1,p_2)=\frac{{E(p_1)-E(p_2)}}{{2\sqrt{E(p_1)E(p_2)}}}$%
\end{tabular}
.  \label{f27}
\end{equation}
Consequences from (\ref{f25}) are expressions for even $[\hat A]$ and odd $\{%
\hat A\}$ parts of the operator of a charge-invariant observable in terms of
its Weyl symbol:
\begin{equation}
\left[ \hat A^{FV}\right] _\alpha {} ^\beta =\frac 1{{(2\pi \hbar )^d}}%
\int\limits_{-\infty }^{+\infty }{\left| {p+\frac P2}\right\rangle
A(p,q)\varepsilon (p+\frac P2,p-\frac P2)\delta _\alpha {} ^\beta \exp\left[-\frac i%
\hbar Pq\right]dPdpdq\left\langle {p-\frac P2}\right| },
\label{f28}
\end{equation}
\begin{equation}
\left\{ \hat A^{FV}\right\} _\alpha {} ^\beta =\frac 1{{(2\pi \hbar )^d}}%
\int\limits_{-\infty }^{+\infty }{\left| {p+\frac P2}\right\rangle
A(p,q)\chi (p+\frac P2,p-\frac P2)\tau _1 {}_\alpha {} ^\beta \exp\left[-\frac i%
\hbar Pq\right]dPdpdq\left\langle {p-\frac P2}\right| .}
\label{f29}
\end{equation}

Matrix elements with eigenvectors of the momentum of the operator
of a charge-invariant variable and its even and odd parts can be
written in the following form:
\begin{equation}
\left\langle {p_1}\right| \hat A^{FV} {}_\alpha {}^\beta \left| {p_2}%
\right\rangle =\frac 1{{(2\pi \hbar )^d}}R_\alpha {}^\beta
(p_1,p_2)\int\limits_{-\infty }^{+\infty }{A(\frac 12(p_1+p_2),q)\exp\left[-\frac i%
\hbar (p_1-p_2)q\right]dq},  \label{f30}
\end{equation}
\begin{equation}
\left\langle {p_1}\right| \left[ \hat A^{FV}\right] _\alpha {} ^\beta \left|
{p_2}\right\rangle =\frac 1{{(2\pi \hbar )^d}}\varepsilon (p_1,p_2)\delta
_\alpha {} ^\beta \int\limits_{-\infty }^{+\infty }{A(\frac 12%
(p_1+p_2),q)\exp\left[-\frac i%
\hbar (p_1-p_2)q\right]dq},  \label{f31}
\end{equation}
\begin{equation}
\left\langle {p_1}\right| \left\{ \hat A^ {FV}\right\} {} _\alpha {} ^\beta
\left| {p_2}\right\rangle =\frac 1{{(2\pi \hbar )^d}}\chi (p_1,p_2) \tau _1
{}_\alpha {} ^\beta \int\limits_{-\infty }^{+\infty }{A(\frac 12%
(p_1+p_2),q)\exp\left[-\frac i%
\hbar (p_1-p_2)q\right]dq}.  \label{f32}
\end{equation}
 Then, as was achieved in Sec.\ref{II}, one can obtain a
formula that reconstructs the Weyl symbol from the operator, with
even and odd parts, in the Feshbach–Villars representation:
\begin{equation}
A(p,q)\delta _\alpha {}^\beta =\sum\limits_{\gamma =\pm 1}{%
\int\limits_{-\infty }^{+\infty }{R^{-1} {}_\gamma {}^\beta (p-\frac P2,p+%
\frac P2)\left\langle {p+\frac P2}\right| \hat A ^{FV} {}_\alpha
{}^\gamma \left| {p-\frac P2}\right\rangle \exp\left[\frac i\hbar
Pq\right]dP}},  \label{f33}
\end{equation}
\begin{equation}
A(p,q)\delta _\alpha {}^\beta =\int\limits_{-\infty }^{+\infty }{\varepsilon
^{-1}(p-\frac P2,p+\frac P2)\left\langle {p+\frac P2}\right| }\left[ \hat A%
^{FV} {}\right] _\alpha {} ^\beta \left| {p-\frac P2}\right\rangle \exp\left[\frac i%
\hbar Pq\right]dP,  \label{f34}
\end{equation}
\begin{equation}
A(p,q)\delta _\alpha {} ^\beta =\sum\limits_{\gamma =\pm 1}{%
\int\limits_{-\infty }^{+\infty }{\chi ^{-1}(p-\frac P2,p+\frac P2) \tau _1
{}_\gamma {} ^\beta \left\langle {p+\frac P2}\right| }}\left\{ \hat A^{FV}
{}\right\} {}_\alpha {}^\gamma \left| {p-\frac P2}\right\rangle \exp\left[\frac i%
\hbar Pq\right]dP.  \label{f35}
\end{equation}

Comparing (\ref{f31}) and (\ref{f32}) we conclude that matrix
elements (integral kernels) of even and odd parts of the operator
of an arbitrary charge-invariant variable are uniquely related to
each other due to the Weyl rule:
\begin{equation}
\left\langle {p_1}\right| \left\{ \hat A^{FV} \right\} _\alpha
{}^\beta
\left| {p_2}\right\rangle =\frac{{E(p_1)-E(p_2)}}{{E(p_1)+E(p_2)}}%
\sum\limits_{\gamma =\pm 1}{\tau _1{}_\gamma {}^\beta \left\langle {p_1}%
\right| }\left[ {{\hat A^{FV}}}\right] {{_\alpha }^\gamma \left| {p_2}%
\right\rangle }.  \label{f36}
\end{equation}
A consequence of this expression is the fact that independent of
position the odd part of an operator is zero. If one uses as
$\hat A$, for example, the scalar potential of an electric field,
expression (\ref{f35}) establishes a quantitative relationship
between effects of motion of a particle in an electric field and
its interaction with a polarizable vacuum (trembling motion,
Zitterbewegung) \cite{b12}.

Consider now time derivative peculiarities of charge-invariant
variables. In the coordinate representation, the matrix-valued
Weyl symbol of such an operator has the form:
\begin{equation}
\partial _t{\check A}=\{\check A,\check H\}_M.  \label{f37}
\end{equation}
The Weyl rule (\ref{f23}) (in the Feshbach - Villars
representation) for this can be written as follows:
\begin{equation}
\partial _t{\hat A^{FV}}=\frac 1{{(2\pi \hbar )^d}}\int\limits_{-\infty
}^{+\infty }{\left| {p+\frac P2}\right\rangle }\check U(p+\frac P2)\{\check A%
(p,q),\check H(p)\}_M\check U^{-1}(p-\frac P2)\exp\left[-\frac
i\hbar Pq\right]dPdpdq\left\langle {p-\frac P2}\right| ,
\label{f38}
\end{equation}
or, in another form:
\begin{equation}
\partial _t{\hat A^{FV}}=\frac 1{{(2\pi \hbar )^d}}\int\limits_{-\infty
}^{+\infty }{\left| {p+\frac P2}\right\rangle \{\check A(p,q),\check G%
(p^{\prime }+\frac P2,p,p^{\prime }-\frac P2)\}_M^{p^{\prime }=p}\exp\left[-\frac i%
\hbar Pq\right]dPdpdq\left\langle {p-\frac P2}\right| },
\label{f39}
\end{equation}
where we have introduced a new matrix-valued variable of three arguments:
\begin{equation}
G_\alpha {}^\beta (p_1,p,p_2)=\frac{{E^2(p)}}{{2\sqrt{E(p_1)E(p_2)}}}(\tau
_3{}_\alpha {}^\beta +i\tau _2{}_\alpha {}^\beta ).  \label{f40}
\end{equation}

If $A(p,q)$ on $q$ linearly depends, formula (\ref{f39}) can be
presented in a particularly simple form:
\begin{equation}
\partial _t{\hat A^{FV}{}_\alpha {}^\beta }=\int\limits_{-\infty }^{+\infty }{%
\left| p\right\rangle \frac{{\partial A(p,q)}}{{\partial q}}\frac{{\partial
E(p)}}{{\partial p}}(\tau _3{}_\alpha {}^\beta +i\tau _2{}_\alpha {}^\beta
)dp\left\langle p\right| }.  \label{f41}
\end{equation}
The time derivative of an operator is a composition of its even
and odd parts, where the odd part has a classical limit because
of the absence of interference terms. As an example one can take
the Newton - Wigner position operator that is even part of the
usual position operator.

In the general case this formula contains in higher orders of
$\hbar$
non-standard terms, since $\check{G} (p_1 ,p,p_2 )$ distinguishes from $E(p)$%
. However, in the classical limit they vanish and this expression
takes the form of (\ref{f41}).

Hence, in the Feshbach - Villars representation, where we
distinguish between solutions with different charge signs, the odd
part of the position results not only in the emergence of the odd
part of operators, but leads to some peculiarities in their even
parts as well. In particular, it appears in the specifics of the
semi-classical limit.

\section{Wigner function and quantum Liouville equation for charge-invariant
variables}

\label{V}

It is easy to see from (\ref{f25}) that it is possible to
introduce the usual Wigner function for the charge invariant
variables in such a way that their average values are determined
by the formula:
\begin{equation}
\bar{A}=\int\limits_{-\infty }^{+\infty }{A(p,q)W(p,q)dpdq}.  \label{f42}
\end{equation}
To show this, we expand the operator of quasi-probability density
in the Feshbach - Villars representation through the eigenvectors
of momentum:
\begin{equation}
\hat W ^{FV} {}_{\alpha } {}^{\beta }(p,q)=\frac{1}{{(2\pi \hbar
)^{d}}}\int\limits_{-\infty }^{+\infty }
{\left| {p+\frac{P}{2}}\right\rangle R_{\alpha } {}^{\beta }(p+\frac{P}{2},p-%
\frac{P}{2})\exp\left[-\frac{i}{\hbar }Pq\right]dP\left\langle
{p-\frac{P}{2}}\right| }. \label{f43}
\end{equation}
The Wigner function is determined as the average value of this
operator over an arbitrary state that contains in the general case
components with both charge signs:
\begin{equation}
W(p,q)=\sum\limits_{\alpha ,\beta =\pm 1}{\left\langle {\psi _{\beta }}%
\right| } \hat W^{FV} {}_{\alpha } {}^{\beta }(p,q)\left| {\psi ^{\alpha }}%
\right\rangle .  \label{f44}
\end{equation}
There are four components in this expression. Two of them are the
average value of the even part of the operator of
quasi-probability density, and the other two are the average
values of the odd part. Let us introduce the symbols:
\begin{equation}
W_{\alpha } {}^{\beta }(p,q)=\left\langle {\psi _{\beta }}\right| \hat W%
^{FV} {}_{\alpha } {}^{\beta }(p,q)\left| {\psi ^{\alpha }}\right\rangle .
\label{f45}
\end{equation}
It should be noted here and below that the object $W_{\alpha }
{}^{\beta }(p,q)$ is not the matrix-valued Wigner function in the
sense of \cite{b20} and Sec.\ref{III} of this work.

Substituting (\ref{f43}) into (\ref{f45}), we obtain for the
Wigner function components following expressions
\begin{equation}
W_\alpha {}^\alpha (p,q)=\frac 1{{(2\pi \hbar )^d}}\int\limits_{-\infty }^{+\infty } {\varepsilon (p+%
\frac P2,p-\frac P2)\psi _\alpha ^{*}(p+\frac P2)\psi ^\alpha (p-\frac P2%
)\exp\left[-\frac i\hbar Pq\right]dP},  \label{f46}
\end{equation}
\begin{equation}
W_\alpha {}^{-\alpha }(p,q)=\frac 1{{(2\pi \hbar )^d}}\int\limits_{-\infty }^{+\infty } {\chi (p+\frac P2%
,p-\frac P2)\psi _\alpha ^{*}(p+\frac P2)\psi ^{-\alpha }(p-\frac
P2)\exp\left[-\frac i\hbar Pq\right]dP}.  \label{f47}
\end{equation}
Even components of the Wigner function (\ref{f46}) correspond to a
charge definite state. The value of odd components (\ref{f47}) for
such a state is zero. The expression (\ref{f46}) differs from
analogous one for a non-relativistic Wigner function and
relativistic one determined using the Newton - Wigner position
operator \cite{b23} by the function $\varepsilon (p_1,p_2)$under
the integral sign (see (\ref{f27})). This function has a specific
feature: its expansion on $p_1$, $p_2$ does not contain square
terms. This means that in the non-relativistic limit expression
(\ref{f46}) coincides with the usual determination of the Wigner
function.

We obtain the evolution equations for every component separately.
The general principle here is the same as in Sec.\ref{III}.
However, instead of equation (\ref{f9}) we shall use the integral
form of Klein -Gordon equation in the Feshbach - Villars
representation \cite{b22}:
\begin{equation}
i\hbar \partial _t{\Psi _\alpha (p)}=\sum\limits_{\beta =\pm
1}{E(p)\tau _3{}^\beta {}_\alpha \Psi _\beta (p)}. \label{f48}
\end{equation}
The following equations can be obtained in a standard way, through
differentiating Wigner function components with respect to time:
\begin{equation}
\partial _t{W_\alpha {}^\alpha (p,q,t)}=\alpha \frac 2\hbar E(p)\sin \{-\frac
\hbar 2\overleftarrow{\partial }_p\overrightarrow{\partial }_q\}W_\alpha
{}^\alpha (p,q,t),  \label{f49}
\end{equation}
\begin{equation}
\partial _t{W_\alpha {}^{-\alpha }(p,q,t)}=i\alpha \frac 2\hbar E(p)\cos \{%
\frac \hbar 2\overleftarrow{\partial }_p\overrightarrow{\partial }%
_q\}W_\alpha {}^{-\alpha }(p,q,t).  \label{f50}
\end{equation}

Nevertheless, the Wigner function components are not independent, i.e. a
specific constraint is imposed on solutions of the system (\ref{f49}),(\ref{f50})%
. To find this one should take the Fourier transform and make the
standard change of variables for every component
(\ref{f46}),(\ref{f47}). As a result, we obtain the following
expressions:
\begin{equation}
\Psi _{+}^{\ast }(p_{1})\Psi ^{+}(p_{2})=\varepsilon
^{-1}(p_{1},p_{2})\int\limits_{-\infty }^{+\infty }{W_{+} {}^{+}(\frac{1}{2}%
(p_{1}+p_{2}),q)\exp\left[\frac{i}{\hbar
}(p_{1}-p_{2})q\right]dq}, \label{f51}
\end{equation}
\begin{equation}
\Psi _{-}^{\ast }(p_{1})\Psi ^{-}(p_{2})=\varepsilon
^{-1}(p_{1},p_{2})\int\limits_{-\infty }^{+\infty }{W_{-} {}^{-}(\frac{1}{2}%
(p_{1}+p_{2}),q)\exp\left[\frac{i}{\hbar
}(p_{1}-p_{2})q\right]dq},  \label{f52}
\end{equation}
\begin{equation}
\Psi _{+}^{\ast }(p_{1})\Psi ^{-}(p_{2})=\chi
^{-1}(p_{1},p_{2})\int\limits_{-\infty }^{+\infty }{W_{+} {}^{-}(\frac{1}{2}%
(p_{1}+p_{2}),q)\exp\left[\frac{i}{\hbar
}(p_{1}-p_{2})q\right]dq},  \label{f53}
\end{equation}
\begin{equation}
\Psi _{-}^{\ast }(p_{1})\Psi ^{+}(p_{2})=\chi
^{-1}(p_{1},p_{2})\int\limits_{-\infty }^{+\infty }{W_{-} {}^{+}(\frac{1}{2}%
(p_{1}+p_{2}),q)\exp\left[\frac{i}{\hbar
}(p_{1}-p_{2})q\right]dq}.  \label{f54}
\end{equation}
Now we divide (\ref{f52}) by (\ref{f53}) and (\ref{f54}) by
(\ref{f51}), and equate the resulting expressions with each other
due to the equality of their left-hand sides. This gives us the
constraint we are looking for:
\begin{equation}
\begin{tabular}{l}
$(E(p_{1})-E(p_{2}))^{2}\int\limits_{-\infty }^{+\infty }{W_{+} {}^{+}(\frac{%
1}{2}(p_{1}+p_{2}),q_{1})W_{-} {}^{-}(\frac{1}{2}(p_{1}+p_{2}),q_{2})\exp\left[%
\frac{i}{\hbar }(q_{1}+q_{2})(p_{1}-p_{2})\right]dq_{1}dq_{2}=}$ \\
$=(E(p_{1})+E(p_{2}))^{2}\int\limits_{-\infty }^{+\infty }{W_{+} {}^{-}(%
\frac{1}{2}(p_{1}+p_{2}),q_{1})W_{-} {}^{+}(\frac{1}{2}%
(p_{1}+p_{2}),q_{2})\exp\left[\frac{i}{\hbar }%
(q_{1}+q_{2})(p_{1}-p_{2})\right]dq_{1}dq_{2}}$%
\end{tabular}
.  \label{f55}
\end{equation}

Equation (\ref{f50}) explicitly contains the imaginary unit, so
the actual question is whether the Wigner function is real. To
test this we consider the complex-conjugate expressions for
(\ref{f46}),(\ref{f47}). After some easy transformations we
obtain the following identities:
\begin{equation}
W^{\ast } {}_{\alpha } {}^{\alpha }(p,q)=W_{\alpha } {}^{\alpha }(p,q),
\label{f56}
\end{equation}
\begin{equation}
W^{\ast } {}_{\alpha } {}^{-\alpha }(p,q)=W_{-\alpha } {}^{\alpha }(p,q).
\label{f57}
\end{equation}
These mean that even components of the Wigner function are real.
Odd components are complex conjugate to each other so that their
sum is real as well.

It is essential that the equation for even components of the
Wigner function coincides with the analogous expression obtained
in \cite{b23} for the formalism where the Newton - Wigner position
operator is used. Hence, the dynamics of quasi-distribution
functions for systems of particles with charges of the same sign
(charge definite states) is identical in both cases.

\section{Statistical properties of the Wigner function for charge-invariant
variables}

\label{VI}

Constraint on the initial conditions of the Wigner function is
the general peculiarity of the approach described here because
equations are identical in both cases (for charge definite
states). In this Section we show how some theorems and properties
differ from their analogues in the usual WWM formalism \cite{b16}
and in an approach where the Newton - Wigner position operator is
used \cite{b23}.

First of all one should note the property of normality. Even part
of the Wigner function (\ref{f46}) is normalized in the whole
phase space, and the integral of the odd part (\ref{f47}) is zero.

Consider now the compatibility of the Wigner function (\ref{f46})
with distributions in the coordinate and momentum spaces for a
single-charge state. For this purpose we integrate (\ref{f46}) by
coordinate. As a result, we obtain the distribution in momentum
space:
\begin{equation}
W_{\alpha } {}^{\alpha }(p)=\Psi _{\alpha }^{\ast }(p)\Psi ^{\alpha }(p).
\label{f58}
\end{equation}
This function always has a definite sign, so that it can be
interpreted as the probability density. One can obtain a more
non-trivial result for distribution in coordinate space. The
result of integrating of (\ref{f46}) by momentum can be given in
the following form:
\begin{equation}
W_{\alpha } {}^{\alpha }(q)=\frac{1}{{(2\pi \hbar )^{d}}}\int\limits_{-%
\infty }^{+\infty }{\Psi _{\alpha }^{\ast }(p_{1})\exp\left[-\frac{i}{\hbar }%
p_{1}q\right]\varepsilon (p_{1},p_{2})\Psi ^{\alpha }(p_{2})}\exp\left[\frac{i}{\hbar }%
p_{2}q\right]dp_{1}dp_{2},  \label{f59}
\end{equation}
or, which is the same,
\begin{equation}
W_{\alpha } {}^{\alpha }(q)=\Psi _{\alpha }^{\ast }(q)\varepsilon (i\hbar
\overleftarrow{\partial }_{q},i\hbar \overrightarrow{\partial }_{q})\Psi
^{\alpha }(q),  \label{f60}
\end{equation}
where $\Psi _{\alpha }(q)$ is the wavefunction in the
representation of the Newton - Wigner coordinate \cite{b12}. It
is obvious, that quasi-distribution, (\ref{f59}) and (\ref{f60}),
is not sign-definite. This property is typical for bosons and
causes difficulties in probability interpretation \cite{b28,b29}.
Nevertheless, formal use of such quasi-probability makes it
possible to calculate average values of the variables dependent
on coordinate.

Average values of variables only depending on momentum do not
differ from similar ones in usual approach. Let us find the
peculiarities of higher moments of coordinate in our case. For
this purpose we integrate (\ref{f59}) with $q^{n}$ by coordinate
and use the identity:
\begin{equation}
q^{n}e^{\frac{i}{\hbar }(p_{2}-p_{1})q}=(-i\hbar \overrightarrow{\partial }%
_{p_{2}})^{n}\exp\left[\frac{i}{\hbar }(p_{2}-p_{1})q\right].
\label{f61}
\end{equation}
After some obvious transformations the result for the $n$-th
moment of the coordinate can be written as follows:
\begin{equation}
\left\langle {q^{n}}\right\rangle =\int\limits_{-\infty }^{+\infty }{\left\{
{\Psi _{\alpha }^{\ast }(p)\left[ {i\hbar \overrightarrow{\partial }_{p}}%
\right] ^{n}\Psi ^{\alpha }(p)\varepsilon (p,p^{\prime })}\right\}
_{p^{\prime }=p}dp}.  \label{f62}
\end{equation}
The first moment (average coordinate) has a value similar to one
in the Newton - Wigner coordinate approach. Differences manifest
themselves in higher moments. As an example, we apply this
expression to second moment of the coordinate:
\begin{equation}
\left\langle {q^{2}}\right\rangle =\int\limits_{-\infty }^{+\infty }{\Psi
_{\alpha }^{\ast }(p)\left[ {i\hbar \overrightarrow{\partial }_{p}}\right]
^{2}\Psi ^{\alpha }(p)dp}-\int\limits_{-\infty }^{+\infty }{\Psi _{\alpha
}^{\ast }(p)\left[ {\frac{{\hbar c^{2}p}}{{2E^{2}(p)}}}\right] ^{2}\Psi
^{\alpha }(p)dp} .  \label{f63}
\end{equation}
As well as the usual part, this formula also contains an
additional term that causes the peculiarities related to
determination of the position operator. For strongly localized
states, it results in formal violation of the uncertainty
relation.

Similar to the case of the usual WWM formalism \cite{b16}, we
prove two criteria that make it possible to select Wigner
functions for pure and mixed states out of the whole set of
functions of the variables $(p,q)$.

\begin{criterion}
Criterion of pure state.

For the functions $W_{\alpha } {}^{\alpha }(p,q)$ and $W_{\alpha }
{}^{-\alpha }(p,q)$ to be even and odd components of the Wigner
function for charge invariant variables, it is necessary and
sufficient that equalities (\ref {f55}),(\ref{f56}),(\ref{f57})
hold true, and the following conditions are satisfied:
\begin{equation}
\frac{{\partial ^{2}}}{{\partial p_{1}\partial p_{2}}}\ln
\int\limits_{-\infty }^{+\infty }{W_{\alpha } {}^{\alpha }(\frac{1}{2}%
(p_{1}+p_{2}),q)\exp\left[\frac{i}{\hbar }(p_{1}-p_{2})q\right]dq=-\frac{{c^{4}p_{1}p_{2}}%
}{{E(p_{1})E(p_{2})(E(p_{1})+E(p_{2}))^{2}}}},  \label{f64}
\end{equation}
\begin{equation}
\frac{{\partial ^{2}}}{{\partial p_{1}\partial p_{2}}}\ln
\int\limits_{-\infty }^{+\infty }{W_{\alpha } {}^{-\alpha }(\frac{1}{2}%
(p_{1}+p_{2}),q)\exp\left[\frac{i}{\hbar }(p_{1}-p_{2})q\right]dq=-\frac{{c^{4}p_{1}p_{2}}%
}{{E(p_{1})E(p_{2})(E(p_{1})-E(p_{2}))^{2}}}}.  \label{f65}
\end{equation}
\end{criterion}

Necessity. Following \cite{b16} we start from an obvious identity:
\begin{equation}
\frac{{\partial ^{2}}}{{\partial p_{1}\partial p_{2}}}\ln \Psi _{\alpha
}^{\ast }(p_{1})\Psi ^{\beta }(p_{2})\equiv 0.  \label{f66}
\end{equation}
Applying it to formulas (51) - (54) one obtains the equalities
\begin{equation}
\frac{{\partial ^{2}}}{{\partial p_{1}\partial p_{2}}}\ln
\int\limits_{-\infty }^{+\infty }{W_{\alpha } {}^{\alpha }(\frac{1}{2}%
(p_{1}+p_{2}),q)\exp\left[\frac{i}{\hbar }(p_{1}-p_{2})q\right]dq=}\frac{{\partial ^{2}}}{%
{\partial p_{1}\partial p_{2}}}\ln \varepsilon (p_{1},p_{2}),  \label{f67}
\end{equation}
\begin{equation}
\frac{{\partial ^{2}}}{{\partial p_{1}\partial p_{2}}}\ln
\int\limits_{-\infty }^{+\infty }{W_{\alpha } {}^{-\alpha }(\frac{1}{2}%
(p_{1}+p_{2}),q)\exp\left[\frac{i}{\hbar }(p_{1}-p_{2})q\right]dq=\frac{{\partial ^{2}}}{{%
\partial p_{1}\partial p_{2}}}\ln \chi (p_{1},p_{2})}.  \label{f68}
\end{equation}
Now, substituting the explicit form of $\varepsilon
(p_{1},p_{2})$ and $\chi (p_{1},p_{2})$from (\ref{f27}), we obtain
(\ref{f64}),(\ref{f65}).

Sufficiency. Let components of the function satisfy conditions (\ref{f64}),%
(\ref{f65}) or, equivalently, (\ref{f67}),(\ref{f68}). So one can
write following conditions:
\begin{equation}
\varepsilon ^{-1}(p_{1},p_{2})\int\limits_{-\infty }^{+\infty }{W_{\alpha }
{}^{\alpha }(\frac{1}{2}(p_{1}+p_{2}),q)\exp\left[\frac{i}{\hbar }(p_{1}-p_{2})q\right]dq=%
}\varphi _{1}(p_{1},\alpha )\varphi _{2}(p_{2},\alpha ),  \label{f69}
\end{equation}
\begin{equation}
\chi ^{-1}(p_{1},p_{2})\int\limits_{-\infty }^{+\infty
}{W_{\alpha } {}^{-\alpha
}(\frac{1}{2}(p_{1}+p_{2}),q)\exp\left[\frac{i}{\hbar
}(p_{1}-p_{2})q\right]dq=}\varphi _{3}(p_{1},\alpha )\varphi
_{4}(p_{2},\alpha ), \label{f70}
\end{equation}
where $\varphi _{i}(p_{k},\alpha )$ are certain functions. Let us
show that they can be chosen in such a way as to be consistent
with each other and present a wavefunction for a scalar charged
particle. From (\ref{f56}),(\ref{f69}) and
(\ref{f57}),(\ref{f70}) it follows that
\begin{equation}
\varphi _{1}^{\ast }(p_{1},\alpha )\varphi _{2}^{\ast }(p_{2},\alpha
)=\varphi _{1}(p_{2},\alpha )\varphi _{2}(p_{1},\alpha ),  \label{f71}
\end{equation}
\begin{equation}
\varphi _{4}^{\ast }(p_{1},\alpha )\varphi _{3}^{\ast }(p_{2},\alpha
)=\varphi _{4}(p_{2},-\alpha )\varphi _{3}(p_{1},-\alpha ).  \label{f72}
\end{equation}
Then one obtains
\begin{equation}
\frac{{\varphi _{1}^{\ast }(p_{1},\alpha )}}{{\varphi _{2}(p_{1},\alpha )}}=%
\frac{{\varphi _{1}(p_{2},\alpha )}}{{\varphi _{2}^{\ast }(p_{2},\alpha )}}%
=const=A,  \label{f73}
\end{equation}
\begin{equation}
\frac{{\varphi _{3}^{\ast }(p_{1},-\alpha )}}{{\varphi _{4}(p_{1},\alpha )}}=%
\frac{{\varphi _{3}(p_{2},\alpha )}}{{\varphi _{4}^{\ast }(p_{2},-\alpha )}}%
=const=B.  \label{f74}
\end{equation}
Hence, one can say that our system is described by the following
wavefunctions:
\begin{equation}
\tilde{\Psi}^{\alpha }(p)=\sqrt{A}\varphi _{2}(p,\alpha ),  \label{f75}
\end{equation}
\begin{equation}
\tilde{\Psi}_{\alpha }^{\ast }(p)=\frac{{\varphi _{1}(p,\alpha )}}{\sqrt{A}};
\label{f76}
\end{equation}
\begin{equation}
\tilde{\tilde{\Psi}}{}^{-\alpha }(p)=\sqrt{B}\varphi _{4}(p,\alpha
), \label{f77}
\end{equation}
\begin{equation}
\tilde{\tilde{\Psi}}{}_{\alpha }^{\ast }(p)=\frac{{\varphi _{3}(p,\alpha )}}{%
\sqrt{B}}.  \label{f78}
\end{equation}
Now one has to prove that wavefunctions presented in such a way
(with one and two tildes) are consistent with each other. For
this purpose, a combination of expressions (\ref{f75}) -
(\ref{f78}) and (\ref{f69}) and (\ref{f70}) is substituted into
condition (\ref{f55}). As a result, we obtain
\begin{equation}
\tilde{\tilde{\Psi}}{}_{+}^{\ast }(p_{1})\tilde{\tilde{\Psi}}{}^{-}(p_{2})\tilde{%
\tilde{\Psi}}{}_{-}^{\ast }(p_{1})\tilde{\tilde{\Psi}}{}^{+}(p_{2})=\tilde{\Psi}%
_{+}^{\ast }(p_{1})\tilde{\Psi}^{-}(p_{2})\tilde{\Psi}_{-}^{\ast }(p_{1})%
\tilde{\Psi}^{+}(p_{2}).  \label{f79}
\end{equation}
This leads to the following equality:
\begin{equation}
\frac{{\tilde{\tilde{\Psi}}{}_{+}^{\ast
}(p_{1})\tilde{\tilde{\Psi}}{}_{-}^{\ast
}(p_{1})}}{{\tilde{\Psi}_{+}^{\ast }(p_{1})\tilde{\Psi}_{-}^{\ast }(p_{1})}}=%
\frac{{\tilde{\Psi}^{-}(p_{2})\tilde{\Psi}^{+}(p_{2})}}{{\tilde{\tilde{\Psi}}{}%
^{-}(p_{2})\tilde{\tilde{\Psi}}{}^{+}(p_{2})}}=const=1.
\label{f80}
\end{equation}
If this condition is fulfilled, one can accept as the wavefunction
\begin{equation}
\Psi _{\alpha }(p)=\tilde{\tilde{\Psi}}_{\alpha }(p).  \label{f81}
\end{equation}
Then condition (\ref{f70}) is satisfied and, taking account of the
fact that components of $\tilde{\Psi}_{\alpha }(p)$ differ from
components of $\Psi _{\alpha }(p)$ by a constant phase, condition
(\ref{f69}) is satisfied as well. Hence, conditions of this
criterion allow us to introduce wavefunction (\ref{f81}). Then,
similar to how it is done in \cite{b16}, one can show that it
satisfies the Klein - Gordon equation in the Feshbach - Villars
representation. Which was to be proved.

This criterion differs from a similar one in the usual WWM
formalism \cite{b16} and in the approach that uses the Newton -
Wigner position operator by difference of the right-hand parts of
conditions (\ref{f64}),(\ref{f65}) from zero. The obvious
consequence from this is the fact that the one-particle state,
where Wigner function would be a joint Gauss distribution by
coordinate and momentum, is impossible in the approach described
here.

Consider another property of Wigner function, true for both pure
and mixed states. For this purpose, we introduce the formula for
the square of the module of the scalar product of state $\left|
\Psi \right\rangle $ and state $\left| \Phi \right\rangle $:
\begin{equation}
\begin{tabular}{l}
$\left| {\left\langle {\Psi |\Phi }\right\rangle }\right| ^{2}=(2\pi \hbar
)^{d}\sum\limits_{\alpha =\pm 1}{{\int\limits_{-\infty }^{+\infty }{W_{\Psi
} {}_{\alpha } {}^{\alpha }(p,q)\varepsilon ^{-2}(p+\frac{\hbar }{{2i}}%
\overleftarrow{\partial }_{q},p+\frac{\hbar }{{2i}}\overrightarrow{\partial }%
_{q})W_{\Phi } {}_{\alpha } {}^{\alpha }(p,q)dpdq}}+}$ \\
\newline $+(2\pi \hbar )^{d}\sum\limits_{\alpha =\pm 1}{\int\limits_{-\infty
}^{+\infty }{W_{\Psi } {}_{\alpha } {}^{-\alpha }(p,q)\chi ^{-2}(p+\frac{%
\hbar }{{2i}}\overleftarrow{\partial }_{q},p+\frac{\hbar }{{2i}}%
\overrightarrow{\partial }_{q})W_{\Phi } {}_{-\alpha } {}^{\alpha }(p,q)dpdq}%
}$%
\end{tabular}
.  \label{f82}
\end{equation}
It is easy to test this by substituting expressions (\ref{f46})
and (\ref{f47}) for the Wigner function components into it. Then,
let $W(p,q)$ presents a mixed state consisting of orthogonal
states described by $W_{n}(p,q)$:
\begin{equation}
W_{\alpha } {}^{\beta }(p,q)=\sum\limits_{n}{s_{n}W_{n}{}_{\alpha
} {}^{\beta }(p,q)}, \label{f83}
\end{equation}
\begin{equation}
\sum\limits_{n}{s_{n}}=1.  \label{f84}
\end{equation}
Taking into account that the sum of $s_{n}$(the probability of
system to be in state $n$) squares is less than unity (a
consequence from (\ref{f84}) ) one obtains for an arbitrary state

\begin{equation}
\begin{array}{l}
\sum\limits_{\alpha =\pm 1} {\int\limits_{-\infty }^{+\infty }{W_{\alpha }
{}^{\alpha }(p,q)\varepsilon ^{-2}(p+\frac{\hbar }{{2i}}\overleftarrow{%
\partial }_{q},p+\frac{\hbar }{{2i}}\overrightarrow{\partial }_{q})W_{\alpha
} {}^{\alpha }(p,q)dpdq}}+ \\
+\sum\limits_{\alpha =\pm 1}{\int\limits_{-\infty }^{+\infty }{W_{\alpha }
{}^{-\alpha }(p,q)\chi ^{-2}(p+\frac{\hbar }{{2i}}\overleftarrow{\partial }%
_{q},p+\frac{\hbar }{{2i}}\overrightarrow{\partial }_{q})W_{-\alpha }
{}^{\alpha }(p,q)dpdq}} \leq \frac{1}{{(2\pi \hbar )^{d}}}
\end{array}
.  \label{f85}
\end{equation}
Moreover, for a pure state this inequality turns into an equality.
Expression (\ref{f85}) can be used as necessary and sufficient
condition for both pure and mixed states.

This condition is written for a state that is the superposition
of states with different charges signs, which is more typical for
the many-particle case. It takes a simpler form for a state where
only one charge sign is realized:
\begin{equation}
\int\limits_{-\infty }^{+\infty }{W_{\alpha } {}^{\alpha }(p,q)\varepsilon
^{-2}(p+\frac{\hbar }{{2i}}\overleftarrow{\partial }_{q},p+\frac{\hbar }{{2i}%
}\overrightarrow{\partial }_{q})W_{\alpha } {}^{\alpha }(p,q)dpdq}\leq \frac{%
1}{{(2\pi \hbar )^{d}}}.  \label{f86}
\end{equation}
Though this inequality does not contain interference terms, it
differs from its non-relativistic analogue due to function
(\ref{f27}) being present. Hence, a dynamical variable that
contains square and higher moments of coordinate can have
peculiarities in many-particle systems that are described by
statistical physics.

\begin{figure}\centerline{\epsfysize=10.0cm \epsfbox{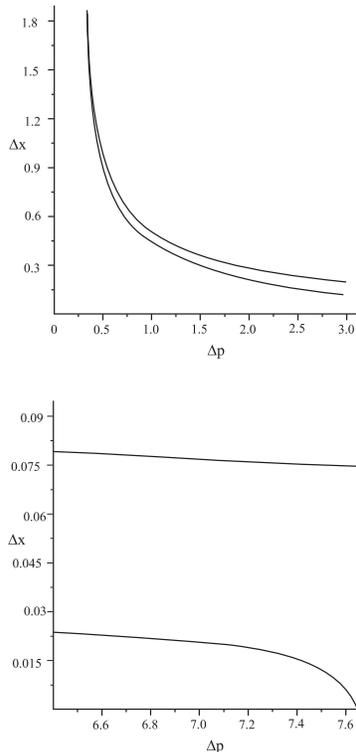}}
\caption{ Dependence of the momentum dispersion on the position
dispersion for the "Gauss" state. Dispersion of the coordinate is
given in $ \lambda _c  = \frac{\hbar }{{mc}}$  units; dispersion
of the momentum is given in $mc$ units. For comparison the curve
$\Delta {\rm X} = \frac{1}{{2\Delta {\rm P}}}$ for usual
uncertainty relation is shown.} \label{fig2}
\end{figure}

The type of influence of the odd part of the position operator on dynamical
variables can be illustrated with the one-particle state that is the Gauss
distribution in momentum space with arbitrary square of dispersion $%
\overline{\Delta p^2}$:
\begin{equation}
\Psi (p)=\frac 1{\sqrt[4]{{2\pi \overline{\Delta p^2}}}}\exp\left[-\frac{{p^2}}{{4%
\overline{\Delta p^2}}}\right]%
{1 \choose 0}
,  \label{f87}
\end{equation}
where we have chosen the natural units of measure, $\hbar =c=m=1$.
The dependence of position dispersion on momentum dispersion is
shown in Fig.\ref{fig2}. Its peculiarity is the formal violation
of uncertainty relation. Under very strong localization, states
even with negative square of dispersion are possible, which was
shown in \cite{b28,b29}. This fact is a property of scalar charged
particles, and localization peculiarities for fermions rather
different.

\section{Conclusions}

\label{VII}

The usual (not Lorentz invariant) Weyl rule makes it possible to
introduce the Wigner function that is not Lorentz invariant, but
all average values calculated with it coincide with ones
calculated with Lorentz invariant wavefunction. This results in
the fact that quantum mechanics in the Wigner formulation contains
with necessity a measuring device frame. In principle, we can
write the evolution equations using only four-dimensional Lorentz
invariant symbols, but it is necessary to introduce a certain
time-like vector for it. It is the four-velocity of the frame in
which wavefunction collapse occurs, relative to the second
(static) observer. It should be noted, that this approach differs
from \cite{b9,b10} where there is the preferred frame. In the
last cases such a frame has a global sense and its introduction is
related to attempts of correct tachyons description and, as a
consequence, to a possible explanation of the instant quantum
correlations (in a relativistic case) from the position of de
Broglie - Bohm quantum mechanics.

Phase space for a scalar charged particle is not only limited by
three couples of the momenta and coordinates . The charge part of
it exists as well. However, in the approach presented here we
leave the operator nature of such variables without modification.
As a result, the matrix-valued Wigner function is the density
matrix in charge space with standard rules of average values
calculation as well.

If we limit our consideration only to such elements of dynamical
algebra that do not depend on variables of the charge space, it
is possible to introduce the usual Wigner function. This object
differs from the Wigner function for a non-relativistic particle
and from the Wigner function in the Newton - Wigner position
operator approach as well. First of all it should be noted that
it contains four components, corresponding to particles,
antiparticles and their interference with each other. Moreover,
even for the one-particle case, when only one component exists,
the definition of the Wigner function differs, as the result of
odd part of the position operator being present, from the usual
one.

This results in non-standard behaviour of some physical variables,
even in the absence of conditions when particles creation is
possible. However, it is not observed  for all physical variables.
For example, energy and number of particles (that are usually
considered in statistical physics) do not show such
peculiarities. Hence, one can expect such effects on the
quadratic and higher moments of coordinate. As an example, one
can take the dispersion that can be interpreted as the real
(physical) size of system that is not limited by external borders.

One can separate two groups of effects that result from this
approach: ones related to interference between particles and
antiparticles, and ones that take place in systems with same
charge signs. Effects of the first group result from presence of
the odd part of the Wigner function, the second ones are due to
the specific function $\varepsilon (p_1,p_2)$ being present in the
even component of the Wigner function. At the one-particle level
it manifests itself, for example, in violation of the uncertainty
relation. Perhaps, such effects can exist in many-particle
systems as well.

Even and odd components of Wigner function in the system of
quantum Liouville equations are not mixed up together. This
results from absence of particles creation from vacuum in the
system. For example, in an electric field the vacuum is not stable
\cite{b30}, and this can be interpreted as a consequence of the
odd part of the position operator being present as well. It
manifests itself in mixing up of different components in the
quantum Liouville equation. There is another situation for the
instant magnetic field: particles are not created and the Wigner
function components are not mixed up \cite{b24}. Nevertheless,
even components of position and momentum of such a system satisfy
not the usual commutation relations but the deformed Heisenberg -
Weyl algebra ones \cite{b31}. These facts can mean that in
external electromagnetic fields the odd part of the position
operator reveals itself to be especially strong.

\end{document}